\documentstyle[12pt]{article}
\begin{document}
\vspace{1.5ex}
\parindent=120mm {AS-ITP-92-44}
\parindent=10pt
\vspace{1.5ex}
\vspace{4ex}
\parskip 8pt
\begin{center}
{\large\bf THE CONSTRAINT FOR THE LOWEST LANDAU LEVEL \\
AND THE EFFECTIVE FIELD THEORY APPROACH \\
FOR THE FRACTIONAL QUANTUM HALL SYSTEM \\}
\end{center}
\vspace{6.4ex}
\centerline{\bf Zhong-Shui Ma$^{a,b}$, Zhao-Bin Su$^a$ }
\vspace{4.5ex}
\centerline{\sf  $^a$ Institute of Theoretical Physics, Academia Sinica}
\centerline{\sf  Beijing 100080, China}\vspace{1ex}
\centerline{\sf  $^b$ Zhejiang Institute of Modern Physics,Zhejiang University
}
\centerline{\sf         Hangzhou 310027, China\footnote{\sf Mailing address.}}
\baselineskip=24pt
\vspace{6ex}
\begin{center}
\begin{minipage}{5in}
\centerline{\bf  ABSTRACT}
\vspace{2ex}

{\it By applying the Dirac quantization method, we build the constraint
that all electrons are in the lowest Landau level into the Chern-Simons
field theory approach for the fractional quantum Hall system and show that
the constraint can be transmuted from hierarchy to hierarchy. For a finite
system, we derive that the action for each hierarchy can be split
into two parts: a surface part provides the action for the edge excitations
while the remaining part is precisely the bulk action for the next hierarchy.
 And the action for the edge could be decoupled from the bulk only
at the hierarchy filling.  }
\end{minipage}
\vspace{1.5ex}
\end{center}

{\raggedright {\sf PACS numbers: 73.20.Dx; 73.50.Jt; 0.3.70.+k\\}}
\vfill
\eject

Since the celebrated discovery of the fractional quantum Hall effect (FQHE)
[1],
a considerable progress [2] has been made in understanding for FQHE
following upon the seminal paper of Laughlin [3]. Motivated
by the analogies between FQHE and superfluidity [4] as well as the
existence of large ring exchanges on a large length scale [5], Girvin and
MacDonald [6] raised a subtle question whether there is an off-diagonal
long range order (ODLRO) in FQHE ground state. By introducing
a 2+1 dimensional bosonization transformation, they did find a sort of ODLRO
for the bosonized Laughlin wave functions. Based on such an observation,
extensive studies on the effective field theory approach for FQHE have
been appeared in the literatures. Among others, the Ginzburg-Landau
Chern-Simons approach (GLCS) [7,8] successively interprets a variety of
properties for the FQHE system from an {\it ab intio} point of view and
the chiral Luttinger liquid approach [9,10] for the edge excitations
exhibits a deep insight for such an interesting system.

Despite the successes for the various effective field theory approaches[7-12],
we still have the question that whether one should build in the constraint
that all electrons are in the lowest Landau level (LLL) from the very
beginning of the Chern-Simons (C-S) field theory approaches for the FQHE. As
we have seen in GLCS, the ``trivial Gaussian fluctuation'' in fact is raised
from the inter-Landau level
degrees of freedom [7]. From a more basic point of view, it is known that FQHE
system is essentially a 1+1 dimensional system. The 1-D nature of the FQHE
system should be a direct consequence of the LLL constraint in the context of
the 2+1 dimensional C-S field theory.  It is also understandable that the 1-D
nature might be rather crucial for the dynamics for FQHE system.

Motivated by the above arguments, in this paper, we build explicitly
the LLL constraint into the C-S field theory description for FQHE
system and show that the constraint can be transmuted from hierarchy to
hierarchy.  After a careful treatment of the partial integrations for the FQHE
action of a finite system, we show further that the action for each hierarchy
can be split into two parts: a surface part provides the action of the edge
excitations while the remaining bulk part is exactly the action for the next
hierarchy. In particular, the surface action for the edge excitations could be
decoupled from the bulk only at the hierarchy filling.  As the primary
consequences, besides the quantization
conditions for FQHE states as well as its hierarchy scheme [13] can be
deduced systematically as usual, we derive
the equations for the fractionally charged vortices which has an interesting
form without any mass scale dependent parameters.  It also does not depend
on whether FQHE has a BCS type of symmetry breaking. This approach
provides further a theoretical background that the vortices (quasi-particles
) of each hierarchy can have only zero effective mass in the context of C-S
field theory, while the ``conventional" vortices often has finite effective
mass contributed from the massive constituting particles.

For a 2-D N-electron system in a strong perpendicular magnetic
field B, if all electrons are in the LLL, it can be described by the
following Lagrangian [5] as

$${\cal L}=-{e \over c}\sum_i {\dot {\bf r}}_i(t)\cdot {\bf A}({\bf r}_i(t))
-\sum_{i<j}V({\bf r}_i-{\bf r}_j)
\eqno{(1)}$$
where ${\bf r}_i(t)$ is the 2-D coordinate for the i-th electron with $i=
1,...,N$, ${\dot {\bf r}}_i(t)={d{\bf r}_i(t)}/ {dt}$, ${\bf A}({\bf r})$
is the vector potential for the uniform applied magnetic field,
$\bigtriangledown
\times {\bf A}=B$ and $V({\bf r}_i-{\bf r}_j)$ is the interaction between
electrons. Because $\partial {\cal L}/ \partial {\dot {\bf r}}_i(t)$,
$i=1,...,N$, are independent of ${\dot {\bf r}}_i(t)$'s and $[{\Pi_\alpha}^i,
{\Pi_\beta}^j]=-i\in_{\alpha \beta}\delta_{ij}\hbar^2 \lambda^{-2}$,
so that we have the
second class constraints [14] as
$\Pi_\alpha^i \equiv p_\alpha^i+(e / c)A_\alpha^i ({\bf r}_i)\doteq 0$,
where $\alpha, \beta =1,2$ is the component index for the 2-D vectors,
$\in_{\alpha \beta}$ are the invariant second rank
antisymmetric tensor, and $\lambda =(\hbar c /eB)^{1 \over 2}$ the magnetic
length. We then have the Hamiltonian for the N-electron in the LLL as
$$H=\sum_{i<j} V({\bf r}_i-{\bf r}_j)
\eqno{(2)}$$
which is associated with the commutation relation $[x_i^\alpha, x_j^\beta ]
=i \in_{\alpha \beta}\delta_{ij}\lambda^2$, or, equivalently, with the
constraint for the N-electron wave function as $\Pi_i\Psi ({\bf r}_1,...,
{\bf r}_N)=0 $,
where $\Pi_i=(\Pi_1^i-i\Pi_2^i )/{\sqrt 2}$. We now introduce further the
second quantization representation for N-electron system described above.
It is straightforward to verify that the second quantized Hamiltonian
has the form as
$$H=V[\Psi^+\Psi -{\bar \rho}]={1 \over 2}\int d^2xd^2x'(\Psi^+(x)
\Psi (x)-{\bar \rho})V({\bf x}-{\bf x}')(\Psi^+ (x')\Psi (x')-{\bar \rho})
\eqno{(3)}$$
where the electron wave field operator $\Psi (x)$ subjected to a
constraint that
$$\Pi {\widehat \Psi}(x)=0
\eqno{(4)}$$
and $\bar \rho $ is the average electron density contributed by the positive
background.

We further
introduce [6-8] the bosonized representation $\Phi (x)$ for the electron wave
field with the help of a C-S field ${\bf a}(x)$. By applying the standard
procedure, the Hamiltonian actually have the same form as eq. (3), $H=V[\Phi^+
\Phi -{\bar \rho}]$ and the LLL constraint becomes ${\tilde \Pi}\Phi (x)
=0 $, in which ${\tilde \Pi }=\partial / \partial z+iA/(a\lambda^2 B) +ia$ with
$z=(x+iy)/{\sqrt 2}$, $A=(A_1-iA_2)/{\sqrt 2}$  and $a =(a_1 -ia_2)/{\sqrt 2}$.
We have also an additional constraint for the C-S field as
$\nabla \times {\bf a}=-2\pi m \Phi^+\Phi $,
with $m$ being an odd integer. Base upon the above discussions, the path
integral representation for the $Z$-generating functional would have the
following form as
$$Z[A]=\int {\cal D}\Phi {\cal D}\Phi^+ {\cal D}a_\mu \delta [{\tilde \Pi}
\Phi ]\delta [\Phi^+{\tilde \Pi}^+]$$
$$\times exp i\{\Phi^+ (i{\partial \over \partial t}-e\varphi -a_0)\Phi
-V[\Phi^+\Phi -{\bar \rho}]-{1 \over 4\pi m }\in_{\mu \nu \lambda}a_\mu
\partial_\nu a_\lambda \}
\eqno{(5)}$$
where the gauge fixing condition is understood involved implicitly. In eq.(5),
$\mu, \nu $, and $\lambda$ run over both the time index $0$ and the
spatial indices 1,2; $\in_{\mu \nu \lambda}$ is the 3-D euclidean antisymmetric
tensor; $a_0$ is the temporal component for the C-S gauge field playing a role
as a $\lambda$-multipler and $\delta[\cdots]$ is the $\delta-$functional.
{}From now on, we would also include an applied
electric field where $\varphi (x)$ is the corresponding scalar potential.
Since now we are in the boson representation, we prefer to introduce the phase
$\theta (x)$ and the electron density $\rho (x)$ for the wave field as the
dynamical variables: $\Phi (x) ={\sqrt {\rho (x)}}exp(i\theta (x))$ in which
$\theta (x) $ could be split into $\theta_r (x)+\theta_s (x) $. $\theta_r
(x)$ is the regular part satisfying
$\in_{\alpha \beta} \partial_\alpha \partial_\beta \theta_r=0$ and
$\theta_s(x)$
the singular part satisfying $\in_{\alpha \beta}\partial_\alpha \partial_\beta
\theta_s=-\rho_s /(2\pi)$.
with $\rho_s$ having the physical intuition as the density for the vortices.
Taking into considerations of the expression for the ${\tilde {\Pi}}$ and
${\tilde {\Pi}}^+ $, as a result of introducing
the $\rho$-$\theta$ representation, the C-S field acquires a pure gauge
term as: $a_\mu
\to a_\mu +\partial _\mu \theta_r $, $\mu =0,1,2$, in the matter part of the
action. Since the action for the C-S gauge field itself is invariant
respect to the local gauge transformation up to a surface term,
we may eliminate the regular part of the phase variable $\theta_r
(x)$ and would come back to the resulting surface term later on. It is then
straightforward to solve the constraint in terms of $\rho $-$\theta $ variable
and further carry out the integration over the C-S field,
we obtain
$$Z[A]=\int {\cal D}\rho {\cal D}\theta_s \delta [{\cal F}[\rho, \rho_s; B]]
expi\{-\rho {\dot \theta}_s -e\rho \varphi +{1 \over 4\pi
m}\in_{\alpha \beta }a_\alpha {\dot a}_\beta -V[\rho-{\bar \rho}]\}
\eqno{(6)}$$ with
$${\cal F}[\rho,\rho_s; B]\equiv {1 \over 2}\bigtriangledown^2 \ln \rho
+{1 \over \lambda^2}-2\pi m\rho -2\pi \rho_s=0
\eqno{(7)}$$
and $a_\alpha $ being now the solution of $\bigtriangledown \times {\bf a}
=-2\pi \rho $ in consistency with
certain gauge fixing constraints. We also notice the term $\lambda ^{-2}$
in eq.(7) could be understood as $(e/ \hbar c )\bigtriangledown \times {\bf
A} $. This is the first result we like to derive, i.e., the
$Z$-generating functional
for FQHE in which the constraint for the LLL has been carefully built.
The LLL constraint not only makes the electrons' kinetic energy disappeared,
but
also manifest itself as a functional relation among $\rho$, $\rho_s$ and $B$:
${\cal F}[\rho, \rho_s; B]=0$.

Since $\in_{\alpha \beta}\partial_\alpha \partial_\beta \theta_s$ can be
nonzero only at certain singular 2+1 dimensional world line, so that the vortex
density should have the expression as $\rho_s (x)=\sum_j q_j \delta^2 ({\bf x}
-{\bf x}_j (t))$ where $q_j=\pm 1$ is the vortex charge and ${\bf x}_j(t)$'s
are the world lines for the j-th vortex. If we take a time derivative to the
vortex density, we obtain immediately a conservation theorem as ${\dot \rho}_s
+\partial_\alpha j_s^\alpha =0$,
where the vortex current has the expression $j_s^\alpha (x)=\sum_j q_j {{\dot
x}
_j}^\alpha (t)\delta^2 (x-x_j (t))$, or, equivalently,
$j_s^\alpha (x)=\in_{\alpha \beta} (\partial_0 \partial_\beta
-\partial_\beta \partial_0)\theta_s /(2\pi)$.

It is straightforward to derive from the $Z$-functional the following
equation[15]
$${1 \over 2}\bigtriangledown ^2 <\ln \rho >-2\pi m <\rho > +{1 \over
\lambda^2}
-2\pi <\rho_v >=0
\eqno{(8)}$$
where $<...>$ is the path integral averaging over the normalized $Z$-generating
functional, i.e., average over the physical ground state. This equation
had been derived directly from the constraint equations for the LLL[16].
What we have here more
is to make its connection to the dynamics more explicit. For a uniform
system with zero vortex, we derive the quantization condition for FQHE
states, $<\rho > \equiv {\bar \rho}=(2\pi m\lambda^2 )^{-1}$, immediately.
For a single vortex, we can draw the conclusion easily from eq.(8)
that it carries a fractional charge of $-q/m $. So this equation can be
interpreted as the equation for the vortices (quasi-particles) of the first
hierarchy and its mean field
approximation can be solved without difficulty.  We notice that, different
from the usual Ginzburg-Landau type description, there is no mass scale
dependent parameters appeared in eq.(8). It also does not depend
conceptually on whether
there is a ``BCS type symmetry breaking" in the FQHE state.

It is known that the term $ \bigtriangledown^2\ln\rho /2$
plays a role to cancel the $\delta$-function like singularities in
$\rho_s$.
Therefore, for sake of convenience, we would ignore the
$\bigtriangledown^2 \ln \rho$ term in the following which should be understood
that in the first quantization representation for the vortices, there is always
a term $-\in_{\alpha \beta}\partial_\beta \ln \rho /2$ associated with
$\partial_\alpha \theta_s$ implicitly, while in its second
quantization representation, such a term can be reasonably ignored.

Now we shall separate the surface part of the action from the bulk part for a
finite FQHE system. Before going into the details, we like to introduce
certain descriptions for the boundary of the system. We imagine the 2-D system
being enclosed by a (spatially) one dimensional boundary $\Gamma$. We further
introduce a displacement vector $\delta {\bf r}$ which is defined formally
along the boundary as
$$\int d^2x (\rho-{\tilde \rho})\equiv -{\tilde \rho}\oint_\Gamma dl n_\alpha
\delta r_\alpha
\eqno{(9)}$$
where $dl$ is the linear integral along the boundary, $n_\alpha$ is the
normal for the boundary being defined always toward outside of the system and
$\tilde \rho $ is certain average electron density to be defined.  If we
take ${\tilde \rho} ={\bar \rho}$, the lefthand side of the equation should
be zero so that we should have $\oint_\Gamma dl n_\alpha \delta r_\alpha =0$.
Consequently, $\delta r_\alpha $ can be interpreted either as the displacement
for the particles (electrons) passing back and forth through the boundary
or as the `` rippling '' displacement for the boundary [10] deviating out-
or inward along the boundary. Obviously, this equation
is valid up to the first order of $\delta {\bf r}$. Comparing with the
constraint equation (8), we find ${\bar \rho}=(2\pi m\lambda^2 )^{-1} -m^{-1}
{\bar \rho}_s $ with ${\bar \rho}_s$ being the average density of vortices.
If we split $\theta_s$ into two parts: $\theta_s =\theta_s^b+\theta_s^c$,
correspondingly, $\rho_s =\rho_s^b+\rho_s^c$. We have the surface part
of vortex density $\rho_s^c=-\in_{\alpha
\beta}\partial_\alpha \partial_\beta \theta_s^c /2\pi $ which is nonzero
only at the boundary and has zero contribution to the ${\bar \rho}_s $ while
the bulk part of vortex density ${\rho_s}^b=-\in_{\alpha \beta}
\partial_\alpha \partial_\beta {\theta_s}^b/2\pi $ with
${\bar \rho}_s={\bar \rho}_s^b$. By applying further eq.(7) to eq.(9),
we have finally
$\delta r_\alpha =- (2\pi m{\bar \rho})^{-1}[\in_{\alpha \beta}\partial_\beta
\theta_s^c]_\Gamma $
up to an arbitrary gauge transformation ${\theta_s}^c \to {\theta_s}^c+
\theta $ where
$\theta $ is a regular function defined along the $\Gamma$: $\oint dl n_\alpha
\partial_\alpha \theta =0$ but not determined yet.
Moreover, it is known that since a finite 2-D FQHE system are practically
in a confining potential, it acquires a chemical potential $\mu$ in such a way
that the Gibbs free energy being minimized in consistency with the spatial
distribution for the electrons. Therefore, the local deviation for the applied
electric potential, $-e\varphi $, from the chemical potential at the boundary
bears the work done by those electron got passed through the boundary, or in
another words, due to the local displacement of the boundary from its
equilibrium configuration. We should then have $(-e\varphi -\mu )|_\Gamma
=-e{\bf E}\cdot \delta {\bf r}|_\Gamma $.

Now for the first term of the action (see eq. (6)): $-\rho {\dot \theta}_s$,
we introduce a dual gauge field $A'_\alpha =a_\alpha /m$ with $\rho =-
\in_{\alpha \beta}\partial_\alpha A'_\beta /(2\pi)$. By applying a partial
integration with
respect to $\partial_\alpha$, we separate a surface term from the bulk action
with a remaining bulk part which can be expressed in terms of $A'_\alpha $
and $j_{s,\alpha}$. On the meanwhile, we substitute $\rho $ by $(2\pi m)^{-1}
[\lambda^{-2}-2\pi {\rho_s}^b +\in_{\alpha \beta}\partial_\alpha \partial_\beta
{\theta_s}^c]$ into the second term of the action: ${\rho (-e \varphi -\mu )}$.
Noticing the constant ${(2\pi m \lambda^2)}^{-1}$
does not contribute to the dynamics , we perform once again a partial
integration with respect to the ``$\partial_\alpha$'' for the last term,
so that we separate one another surface term from the bulk action. Taking into
account of the above considerations, and further utilizing the expression
for $\delta r_\alpha$,
we obtain an interesting form for the $Z$-functional as
$$Z=\int {\cal D}\theta_s^b{\cal D}\theta_s^c \int {\cal D}\rho
\delta [{\cal F}[\rho, \rho_s^b+\rho_s^c; B]]$$
$$\times expi\{{\bf j}_s^b \cdot {\bf A}'-{e \over m}\rho_s^b\varphi
-{m \over 4\pi}\in_{\alpha \beta}A'_\alpha {\dot A'}_\beta -V-I_\Gamma [\rho,
 \theta_s^s]\}
\eqno{(10)}$$
with the surface action
$$
I_\Gamma [\rho, \theta_s^c]={1 \over 2\pi m}\int dt \oint
dl \{ (-n_\alpha  \in_{\alpha \beta }\partial_\beta \theta_s ){\dot \theta}_s
-{\tilde v}_D (  n_\alpha
\in_{\alpha \beta} \partial_\beta \theta_s^c)^2\}
\eqno{(11)}
$$
In eq.(11), we have assumed the applied electric field is paralleled to the
normal
on the boundary and ${\tilde v}_D=v_D/(1-2\pi\lambda^2{\bar \rho}_s)$ with
$v_D =cE/B $. For a finite system, eq.(10) means that
the action for FQHE system can be divided into a bulk part
and a surface part. The
bulk part has the intuition that the vortices moves in a dual gauge field
${\bf A}'$ and bears a fractional
statistics $1/ m$ with fractional charge $-qe/ m$. If we solve ${A'}_\alpha$
in terms of $\rho $, then applying eq.(7), we recover easily the form
derived in [7]. Moreover, when the system is exactly in a FQHE state of
the first hierarchy, $i.e.$ ${\rho_s}^b=0$, we then have $\theta_s=
{\theta_s}^c$. As an interesting result, the surface action $I_\Gamma
[\rho, {\theta_s}^c]$ will decouple from the bulk and describe an
ensemble of independent propagating edge excitations in a form
of a chiral boson
action proposed in [9,10]. If we perform a gauge
transformation to
the whole action in eq.(10), it would also produce a surface term which may
cancel the surface term left previously.

As we have noticed above,
 $\theta_s (x) $ has only the isolated singularities in the 2-D plane, so that
${\cal D} \theta_s$ integrates over only the space-time propagation of those
singularities: the coordinate of vortices. Therefore
$$\int {\cal D}\theta_s^b expi\{{\bf j}_s^b\cdot {\bf A}'\}
=\sum_{N=1}^\infty \int \prod_{j=1}^N {\cal D}{\bf r}'_j (t) exp i \{
\sum_j {\dot {\bf r}}'_j (t)\cdot {\bf A}' ({\bf r}'_j (t))\}
\eqno{(12)}$$
where ${\bf r}'_j (t) $ is the coordinate for the j-th vortex. For sake of
convenience, we assume in eq.(12) and the following that $q_j=+1$.
This identity
makes it explicit that the bulk action for the vortices
in eq.(12) is essentially in a first quantization representation. Moreover,
it becomes clear also that such an action has only first order time derivative
of the vortex coordinates but no second order. Once again we have a system
of vortices with `` zero kinetic energy '' which should be described again
by a second class constraint. Now the bulk action for the vortices
has a form almost the same as the original action for the electrons, eq.(1).
We can then run the same procedure as for
the electron case, $i.e.$, the procedure from eq.(1) to eq.(11). But there
are still certain delicate differences which should be carefully treated as the
following: (i) Now, for the bulk action for the vortices, we have a vector
potential ${\bf A}'$ playing a similar role as the vector potential for
the magnetic field in electrons case but having a curl $\bigtriangledown
\times {\bf A}'=-2\pi \rho $; (ii) In the application of
the Dirac quantization to the vortices in the first quantization
representation,
we need the condition $[{\Pi'_\alpha}^i, {\Pi'_\beta}^j ]=-2\pi \in_{\alpha
\beta}
\rho \not= 0 $ to be satisfied, where $\Pi^{'i}_{\alpha}$ has the same form
as $\Pi^i_{\alpha}$ with the corresponding quantities substituted by those
for the vortices. Since $\rho $ could be zero (or singular)
only at the isolated
locations for the vortices, in the spirit of long wave length approximation,
we may reasonably take the approximation as $\rho >0$ (finite).
In fact, these singular
behaviors at the vortex locations will disappear after its second quantization
procedure being completed; (iii) Corresponding to the original C-S gauge
field with the
statistical index of odd integers $m$, we now introduce a C-S gauge field
$a'_\mu$ with the statistical index of even integers $2p$, because
the world lines for the vortex ``particles'' are originated from the
singularities of the phase field $\theta_s$ of the bosonized electrons.

Taking into all the above considerations, following almost exactly the same
procedure as those for the electrons, the $Z$-functional can be transformed
into the following form as
$$\int {\cal D}\rho_s {\cal D}\theta'_s \delta [{\cal F}'[\rho_s,{\rho'}_s;B]]
$$
$$\times expi\{-\rho_s {\dot \theta}'_s +{e \over m}\rho_s \varphi -{\kappa
\over {16p^2\pi}}\in_{\alpha \beta}{a'}_\alpha {\dot a'}_\beta
-V[\rho_s]+I_\Gamma
[\rho, \theta_s]\}
\eqno{(13)}$$
with $\kappa =m^{-1}+2p$ and
$${\cal F}'[\rho_s,{ \rho'}_s; B]={1 \over 2}\bigtriangledown^2\ln \rho_s -{1
\over m\lambda^2} +2\pi \rho_s \kappa +\in_{\alpha \beta}\partial_
\alpha \partial_\beta \theta'_s
\eqno{(14)}$$
where $\rho_s$, the density of the vortices, is now in a second quantization
representation, i.e., the modulus of the vortex wave field, and $\theta'_s$
is the singular part for the phase field which describes the isolated
``vortices'' $\rho'_s$ for the next higher hierarchy. In deriving eq.(13),
we have carry out the path integral for ${\cal D}\rho $ so that the constraint
equation (7), ${\cal F}[\rho, \rho_s; B]=0 $, is understood being always
satisfied. Here we notice further that the application of the Dirac's
quantization theory for the constrained systems to the overall space-time
propagation of the vortices in form of eq.(12) provides a field-theoretical
background that the vortices (quasi-particles) in FQHE have only zero effective
mass while the ``conventional"
vortices often have finite effective mass contributed by the massive
constituting particles.

For homogeneous system with ${\bar \rho'}_s$
equals to zero, we have a condensate for both electrons
and vortices. Then
the constraints ${\cal F}[\rho, \rho_s]=0$ and ${\cal F}'[\rho_s,
{\rho'}_s]=0$ results the second hierarchy with a filling $\nu=(m+
(2p)^{-1} )^{-1}$[11].
For systems having isolated ``vortices'' in the sense of the second hierarchy,
${\cal F}'[\rho_s, { \rho'}_s]=0$ provides the corresponding
``vortex''
equation. If we perform a similar procedure to split further the surface
and bulk parts for the bulk action in eq.(13), we derive the corresponding
edge
excitations of the second hierarchy as well as bulk action for the vortex
of the
third hierarchy in which the ``vortex'' current would couple to a dual
C-S gauge field as ${\bf j'}_s^b\cdot {\bf A}''$ with a C-S action
$\kappa \in_{\alpha \beta}A''_\alpha {\dot A}''_\beta /(4\pi )$.
Now it is sufficiently convincing that by repeating the procedure
developed above, we present a full dynamical description for both infinite and
finite FQH systems.  The action incorporated with the constraint can be
transformed from hierarchy to hierarchy in an almost universal form and has
distinguished features which have not been seriously exploited before.

\bigskip
\bigskip

{\raggedright{\large \bf Acknowledgement\\}}
\bigskip

One of the authors (Z.B.S.) would like to thank Profs. L.N.Chang, D.H. Lee, B.
Sakita, S.C. Zhang for useful discussions, especially he likes to
thank to B.Sakita for his kind advisement and encouragement. He would like
to acknowledge also the warm hospitality received from ICTP where the final
version of this work was completed. This work is partially
supported by the NSFC, ITP-CAS and the CCAST.

\bigskip
\bigskip
{\raggedright {\large \bf References\\}}
\bigskip
\begin{enumerate}

\item {D.C. Tsui, H.L. Stormer, A.C. Gossard, {\it Phys. Rev. Lett.}{\bf 48},
1559(1982)}
\item {R. Prange, S.M. Girvin, {\it The Quantum Hall Effect} Springer
Verlag, (1990)}
\item {R.B. Laughlin, {\it Phys. Rev. Lett.}{\bf 50}, 1395(1983)}
\item {S.M. Girvin, A.H. MacDonald, P.M. Platzman, {\it Phys. Rev. Lett.}
 {\bf 54}, 581(1985); {\it Phys. Rev.} {\bf B33},2481(1986)}
\item {S. Kivelson, C. Kallin, D.P. Arovas, J.R. Schrieffer, {\it Phys. Rev.
Lett.}{\bf 56}, 873(1986); G. Baskaran, {\it Phys. Rev. Lett.}{\bf 56},
2716(1986)}
\item {S.M. Girvin, A.H. MacDonald, {\it Phys. Rev. Lett.}{\bf 56},1252(1987)}
\item {S.C. Zhang, H. Hansson, S. Kivelson, {\it Phys. Rev. Lett.}{\bf 62},
82(1989); {\bf 62}, 980(1989); D.H. Lee, S.C. Zhang, {\it Phys. Rev. Lett.}
{\bf 66},
1220(1991); D.H. Lee, {\it Int. J. Mod. Phys.}{\bf B5}, 1695(1991);
S.C. Zhang, {\it Int. J. Mod. Phys.} {\bf B6}, 25(1992) }
\item {N. Read, {\it Phys. Rev. Lett.}{\bf 62}, 86(1989)}
\item {X.G. Wen, {\it Phys. Rev.}{\bf B41}, 12838(1990); {\it Phys. Rev. Lett.}
{\bf 64}, 2206(1990); B. Blok, X.G. Wen, {\it Phys. Rev.}{\bf B42}, 8133(1990);
D.H. Lee, X.G. Wen, {\it Phys. Rev. Lett.}{\bf 66}, 1765(1991); X.G. Wen,
{\it Int. J. Mod. Phys. }{\bf B6}, 1711 (1992) }
\item { M. Stone, {\it Ann. Phys.} (N.Y.) {\bf 207}, 38(1991)}
\item{ X.G. Wen, A. Zee,{\it Phys. Rev.} {\bf B44}, 274 (1991); X.G. Wen,
A. Zee, {\it Phys. Rev. Lett.} {\bf 69}, 953(1992).}
\item{ A. Lopez, E, Fradkin, {\it Phys. Rev.} {\bf B44}, 5246(1991).}
\item {F.D.M. Haldane, {\it Phys. Rev. Lett.}{\bf 51}, 605(1983); B.I.
Halperin, {\it Phys. Rev. Lett.}{\bf 52}, 1583(1984)}
\item {P.A.M. Dirac, {\it Lectures on Quantum Mechanics}, Belfer Graduate
School of Science, Yeshiva University, New York, (1964)}
\item {R.Jackiw and So-Young Pi derived a similar equation but without the
applied magnetic field, {\it Phys. Rev. Lett.} {\bf 64}, 2969(1990).}
\item {B. Sakita, D.N. Sheng, Z.B. Su, {\it Phys. Rev.}{\bf B44}, 11510(1991)}
\end{enumerate}
\end{document}